\documentclass{vgtc}                          %

\graphicspath{{figures/}{pictures/}{images/}{./}} %

\usepackage{times}                     %

\usepackage{tabu}                      %
\usepackage{booktabs}                  %
\usepackage{lipsum}                    %
\usepackage{mwe}                       %

\usepackage{mathptmx}                  %

\onlineid{0}

\vgtccategory{Research}

\vgtcinsertpkg

\usepackage{xspace}
\usepackage{minted}

\makeatletter
\DeclareRobustCommand\onedot{\futurelet\@let@token\@onedot}
\def\@onedot{\ifx\@let@token.\else.\null\fi\xspace}

\def\eg{\emph{e.g}\onedot} 
\def\ie{\emph{i.e}\onedot}

\makeatother

\title{Pluot: Towards `write once, run everywhere' visualization software}

\author{Mark S. Keller\thanks{e-mail: mark\_keller@hms.harvard.edu}\\ %
        \scriptsize Harvard Medical School %
\and Nils Gehlenborg\thanks{e-mail: nils@hms.harvard.edu}\\ %
     \scriptsize Harvard Medical School %
}

\teaser{
  \centering
  \includegraphics[width=\linewidth,trim={4cm 4cm 4cm 4cm},clip]{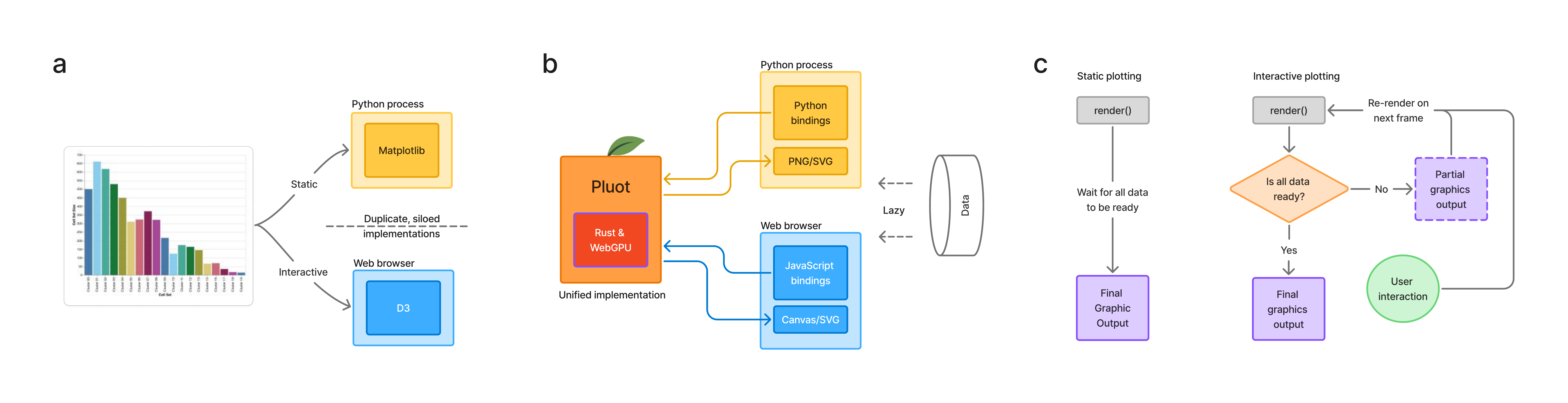}
  \caption{Pluot architecture. a) The status quo. Visualization developers often use different toolkits depending on whether a plot must be static (\eg, for a paper figure) versus interactive (\eg, for a dashboard). b) The Pluot architecture enables developers to implement a visualization using a single toolkit in Rust. Bindings to Rust plotting functions can be generated for other languages such as Python and JavaScript. Pluot scales to out-of-core dataset sizes by performing partial reads of data.
  c) Pluot supports both static and interactive plotting scenarios by treating interactive visualizations as multiple static frames. 
  }
  \label{fig:teaser}
}

\abstract{
    Tools used for implementing visualization software systems can generally be divided into camps such as static versus interactive and desktop versus web-based.
    We contribute Pluot, an architecture that bridges these divides, enabling a single software implementation of a visualization to be used regardless of the target level of interactivity or computing environment.
    With Pluot, a visualization developer implements a given visualization rendering function once, using the Rust programming language.
    Then, bindings to the Rust program can be generated to enable reproducible execution of the rendering function from other languages, such as Python or JavaScript.
    Pluot can render visualizations to bitmap or vector graphics format, bridging gaps between interactive performance and publication-quality figure creation.
    The software is available at \url{https://pluot.dev}.
} %


\begin{document}

\firstsection{Introduction}

\maketitle

Visualization software is currently confined to separate ecosystems, making it challenging to reuse code across ecosystem boundaries.
Visualization developers are often forced to make rigid choices
(\eg, using Matplotlib in Python vs. D3 on the web), impacting interoperability of the resulting software.
Bridging ecosystem boundaries involves additional development effort and complicates maintenance.
Current solutions include porting software from one ecosystem to another, or relying on server-side infrastructure or browser automation tools.
These latter solutions can have high overhead, impacting scalability and performance.
What if, instead, a developer could implement a visualization once, supporting both static and interactive targets, both bitmap and vector graphics representations, and straightforward execution from multiple computing environments?

We present Pluot, a software architecture that unifies visualization development across platforms.
With Pluot, visualization rendering functions are written using Rust.
The Rust program can be compiled to run on Windows, Mac, Linux, or the web, enabling plotting functions to be executed from multiple programming languages using cross-language bindings.

We demonstrate that this approach streamlines the development of visualizations of large-scale datasets encompassing millions of data points.
Pluot supports key interaction types including zooming, panning, hovering, and clicking.
We demonstrate compatibility with coordinated multiple views techniques and faceted designs.  %
Finally, we show that support for both bitmap and vector graphics outputs
simultaneously enables interactive tool development and creation of publication-ready figures.

\section{Design Goals}
\label{sec:design_goals}

We had several design goals based on our experiences implementing static and interactive visualizations in web and non-web contexts
\cite{manz2022viv,lyi2023cistrome,keller2024vitessce}.

\paragraph{DG1. Unified plotting function implementation.} The developer should be able to implement visualization-rendering code once, in a single programming language.
The same code should be able to be reused for static and interactive plotting.
It should be possible to execute this code from multiple contexts, such as a command-line tool, a Python script, a Jupyter notebook, or a standalone web page.
A web browser (headless or otherwise) should not be required to execute the plotting function, as this adds overhead.

\paragraph{DG2. Interactive performance.} Performance should be high enough that plot rendering can be performed on animation frames or on user interactions such as zooming and panning, even for large datasets.
Additional interaction types such as click events, lasso selections, and tooltips should be supported.

\paragraph{DG3. Scalability to out-of-core dataset sizes.} Scientific datasets often comprise hundreds of thousands or millions of data points.
However, based on parameters such as zoom level or a filtering expression, the plotting function may not need to load all data points in order to render the visualization.
While the caller can always subset the data, this complicates implementation of interactive visualizations and duplicates effort (\eg, filtering logic in Python cannot be easily reused in JavaScript).
Instead, it should be possible for the plotting function to `pull' the required subset of data points based on the plotting parameters (\ie, perform partial reads of a given dataset).

\paragraph{DG4. Bitmap and vector output.} It should be possible to return a bitmap (\ie, raster image) or vector (\eg, SVG string) graphics output for the same plot.
Bitmap outputs can benefit from GPU acceleration to achieve high performance (DG2).
Meanwhile, vector outputs can be used for publication-quality figures and accessibility of plot elements. %

\paragraph{DG5. Decoupled from CMV implementation.} To ensure modularity of visualization software, it should remain agnostic to any particular coordinated multiple views (CMV) implementation~\cite{keller2024usecoordination}.
Rather, each plotting function should be concerned with rendering one plot, with multiple views or faceting delegated to the parent software module.
The developer should be able to use their favorite state management approach to implement CMV.

\paragraph{DG6. Decoupled from UI framework.} Interactive plotting functions should not be coupled to a particular UI or window management system.
Rather, interactivity and window management should be delegated to the parent software module, allowing developers to use their favorite UI development framework.
Plotting functions should primarily be concerned with rendering static graphics outputs, reducing the divergence between static and interactive code paths and simplifying the developer mental model.

\section{Related Work}

A 2024 report titled, ``The Moving Target of Visualization Software,'' notes that the combination of WebGPU and WebAssembly is, ``...seen as a promising avenue for `write once, run everywhere' visualization applications while simplifying their development,''~\cite{gillmann2024moving}.
Despite some prior works using WebGPU and WebAssembly for data visualization~\cite{usher2020interactive,herzberger2023residency}, these have not been designed with the `run everywhere' intention: they do not provide bindings to render visualizations from multiple programming languages.

\subsection{Reuse of Web-Based Visualization Software}

Given the large ecosystem of web-based visualization tools, there are plenty of prior efforts to reuse code from the web ecosystem in non-web contexts to avoid reinvention of the wheel.
One approach is to render a webpage that executes visualization-rendering code, and then convert the page to an image (\eg, screenshot with a browser automation tool).
This is a heavyweight approach that can be slow and couples the parent application to a web browser and browser automation tools.
A more lightweight approach is to execute the visualization-rendering code using a ``headless'' JavaScript runtime (\eg, NodeJS) called from the parent application.
This is the approach currently used by vl-convert~\cite{mease2022introducingvlconvert} in the Vega-Lite ecosystem~\cite{satyanarayan2017vegalite}.
However, it is not always possible to run code that was originally designed to run in a web browser in a headless JavaScript runtime, as these do not have complete feature parity.

To reuse web-based software within desktop GUI applications, Electron or simpler ``web views'' can be used.
The ManiVault~\cite{vieth2023manivault} visualization framework enables developers to implement plugins in JavaScript, resulting in visualizations that are rendered to web views within the GUI.
However, ManiVault itself is a desktop application that must be installed on a Windows, MacOS, or Linux platform prior to usage, preventing its usage \textit{within} a web browser or web application.

In the specific case of reuse of web-based software in Python notebook environments, AnyWidget~\cite{manz2024anywidget} streamlines the process of rendering web-based visualizations in these environments.
For example, Jupyter-Scatter~\cite{lekschas2024jupyter} uses AnyWidget to wrap the Regl-Scatterplot library, enabling scalable interactive scatterplot rendering within Jupyter notebooks.
The AnyWidget conceptual approach can be used in other literate programming environments that support web-based outputs, such as htmlwidgets in RMarkdown documents~\cite{githubAnyhtmlwidget}.
AnyWidget is an essential piece of infrastructure that enables visualization developers to reuse web-based software to build notebook widgets, but these widgets and notebook environments still ultimately require a web browser for rendering.

\subsection{Reuse of Non-Web Visualization Software}

Attempts have also been made to reuse non-web visualization software in web-based contexts.
Pyodide~\cite{pyodide2021} uses WebAssembly to enable running Python code in the web browser (Figure~\ref{fig:alternative_approaches}b).
This approach has overhead, requiring the user to download a WASM bundle containing the compiled Python interpreter with many core packages (currently 6.4 MB), plus additional file downloads for each imported Python package.
The Pyodide developers currently report that it runs code 3x to 5x slower than native Python~\cite{pyodide2021}.
In the R ecosystem, WebR is a similar effort, with similar characteristics (initial WASM bundle size, plus additional downloads for imported packages)~\cite{Stagg_webR_The_statistical_2023}.
Thus, Pyodide and WebR can be used to render visualizations in the web browser using popular packages such as Matplotlib or Ggplot2, but their overhead hinder their scalability.

There are also commercial software solutions such as SciChart~\cite{websiteScichart} which implements visualization rendering using C++, with bindings for usage from other programming languages and platforms.
The company provides WebAssembly bindings for web-based usage, as well as bindings for chart rendering in mobile apps developed using Swift/Objective-C (for iOS) and Java/Kotlin (for Android).
SciChart uses DirectX for GPU-accelerated rendering on Windows.
While SciChart has many advanced features, the core software is closed-source, does not support vector graphics export (DG4), and does not provide bindings to languages such as Python or R, excluding many data science use cases.

Client-server architectures can also bridge this gap.
For example, NVIDIA IndeX~\cite{schneider2021nvidia} performs rendering on a remote high-performance compute cluster, with resulting frames streamed to the web browser.
While this approach enables the outputs of non-web-based rendering software to be viewed in a web application, it comes with the complexities and costs of server-side infrastructure.
The usage of Pluot language bindings can be viewed as a local, embedded analog of the remote rendering paradigm: pixels are transmitted over the language boundary (\eg, from Rust to Python) rather than over the network.

\subsubsection{Reuse of Data Processing Logic}
\label{sec:reuse-of-data-processing-logic}

Several recent visualization research projects take advantage of WebAssembly to reuse data processing logic in both web and non-web contexts, as opposed to the visualization rendering logic.
Mosaic~\cite{heer2023mosaic} uses DuckDB-Wasm~\cite{kohn2022duckdbwasm} to accelerate data aggregation and transformation operations.  %
ITK-Wasm~\cite{mccormick2024itkwasm} provides medical image analysis algorithms that have been compiled to WebAssembly.
In the bioinformatics domain, Kana uses single-cell data analysis algorithms that have been compiled to WebAssembly~\cite{lun2023powering}.
Yet these projects all still depend on running code in a web browser or JavaScript runtime to render the processed data to a visualization.

\subsection{Specification-Based Visualization Approaches}
\label{sec:related-work-grammars}

Rendering a visualization to an intermediate grammar-based representation (\ie, a specification) enables interoperability and reproducibility by decoupling the visualization encoding semantics from any software rendering implementation.
Theoretically, the specification-based approach enables implementation of multiple alternative renderers (\eg, in different programming languages).
However, in practice, current widely used visualization grammars depend on a single renderer implemented using web technologies~\cite{satyanarayan2017vegalite,vanderplas2018altair,lyi2022gosling,manz2023gos}.
Therefore, specification-based approaches to achieve reproducibility should be viewed as complementary to a software architecture that enables a single shared rendering implementation that can be reused across computational environments.

\subsection{Usage of Rust for Scientific Computing}

Since being released, the Rust programming language has been increasingly adopted for use cases in scientific computing~\cite{bitar2024rusta}.
In bioinformatics, the libraries Rust-Bio, SingleRust, BigTools, and Polars-Bio implement domain-specific data structures and algorithms in Rust~\cite{koster2016rustbio,diks2025singlerusta,huey2024bigtools,wiewiorka2025polarsbiofast}.
Nature discussed the uptake in a 2020 technology feature titled, ``Why scientists are turning to Rust''~\cite{perkel2020why}.

\subsubsection{Usage of Rust for Data Visualization}

We searched for existing Rust libraries (``crates'') for data visualization\footnote{\url{https://github.com/keller-mark/awesome-rust-vis}}. %
These crates can be subdivided into those in which the visualization logic is implemented in Rust (``pure Rust''), and those which serve as bindings to visualization rendering logic implemented in a different (non-Rust) language.
Among the pure Rust crates, many are coupled to a particular UI framework (DG6).
We did not find an existing crate that satisfied all six of our design goals (Section~\ref{sec:design_goals}).

\section{The Pluot Architecture}

Pluot consists of a core Rust library that performs visualization rendering.
The Pluot library is designed to be compatible with WebAssembly compilation, in addition to desktop operating system targets.
The WebAssembly binary size is currently less than 5MB, making it feasible to load over a network when integrated into web applications.
We generate bindings to the Rust library for other languages, including Python and JavaScript, enabling users to call the visualization rendering functions from these ecosystems.

\subsection{Headless plotting}

The visualization rendering logic in Pluot's Rust core is decoupled from any particular windowing or GUI system (DG6).
Instead, the plot rendering functions return bytes representing either pixels (in the bitmap case) or SVG node strings (in the vector case).
The mechanism used to display these bytes on the screen is left to the caller of plot rendering function.
This makes it easy to integrate Pluot into other systems or to build tools such as command-line interfaces (Section~\ref{sec:cli-gui-http}).
Further, the developer mental model is simplified due to the stateless, functional nature of this approach.

\subsubsection{Language bindings and interactive adapters}
\label{sec:language_bindings_and_adapters}

Within the libraries that contain programming language bindings for non-Rust usage, we provide helper functions to display the returned bytes in static and interactive outputs.
To support interactivity, we provide an \textit{interactive adapter} that handles 2D and 3D camera controls in web environments~\cite{github2dcamera,github3dcamera}.
The interactive adapter additionally handles interaction types such as click events, tooltips triggered by mouse hover events.
The adapter can optionally handle resize events to support responsive layouts and can display loading indicators during data fetching operations.

\subsection{Layer-based API}

Pluot exposes a layer-based API for plotting that simultaneously enables low-level grammar-of-graphics-style declarative usage and high-level chart-type-centric usage.
Developers can define new layers through composition of existing layers (\eg, an \texttt{AxisLayer} combining \texttt{TextLayer} and \texttt{LineLayer} as sub-layers).
Alternatively, developers can define new layer primitives from scratch with full control over the WebGPU shaders, buffers, textures, and render pipeline.

\subsection{Data loading}
Pluot supports asynchronous data loading, enabling data to be loaded via the network or other protocols.
Layers do not need to accept all data up-front, as they can dynamically load data prior to rendering by defining a \texttt{prepare} function.
The \texttt{prepare} function is executed on every rendered frame, enabling data loading to depend on parameters such as zoom level (\eg, to support multi-scale/pyramidal image rendering).
Data can be cached using a \texttt{use\_memo} function inspired by React, preventing duplication of expensive calculations, long network roundtrips, or additional language binding roundtrips~\cite{githubReact}.

We provide a set of layers that load data dynamically via the Zarr array storage format using the Zarrs Rust library~\cite{githubZarrs}.
Zarr stores can be ``registered'' via the programming language bindings that we provide, allowing array data to be requested and passed over the language boundary using the Zarr store interface (Section~\ref{sec:language_bindings_and_adapters}).
For example, when plotting a subset of data from a large Numpy array in Python, we do not pass any array bytes to Rust, but rather we specify a Zarr store and a key at which the array is located.
Our Rust \texttt{prepare} function can then request a subset of data via Zarr's slicing mechanism, which internally will call a Python function that returns the subset of array bytes of interest.
While we find that Zarr's Store abstraction fits this pattern nicely, a layer developer could implement their own \texttt{prepare} function that loads data from an alternative format.

\subsection{Compute shaders}

WebGPU enables developers to implement compute pipelines (in addition to render pipelines), supporting general purpose (GPGPU) operations.
Pluot's layer-based API enables compute pipelines to be executed during the \texttt{layer.prepare} stage, so that the \texttt{layer.draw} stage can render compute shader outputs.
When a GPU context is not available, the developer can implement CPU-based fallbacks.

\begin{figure}[tb]
 \centering %
 \includegraphics[width=\columnwidth]{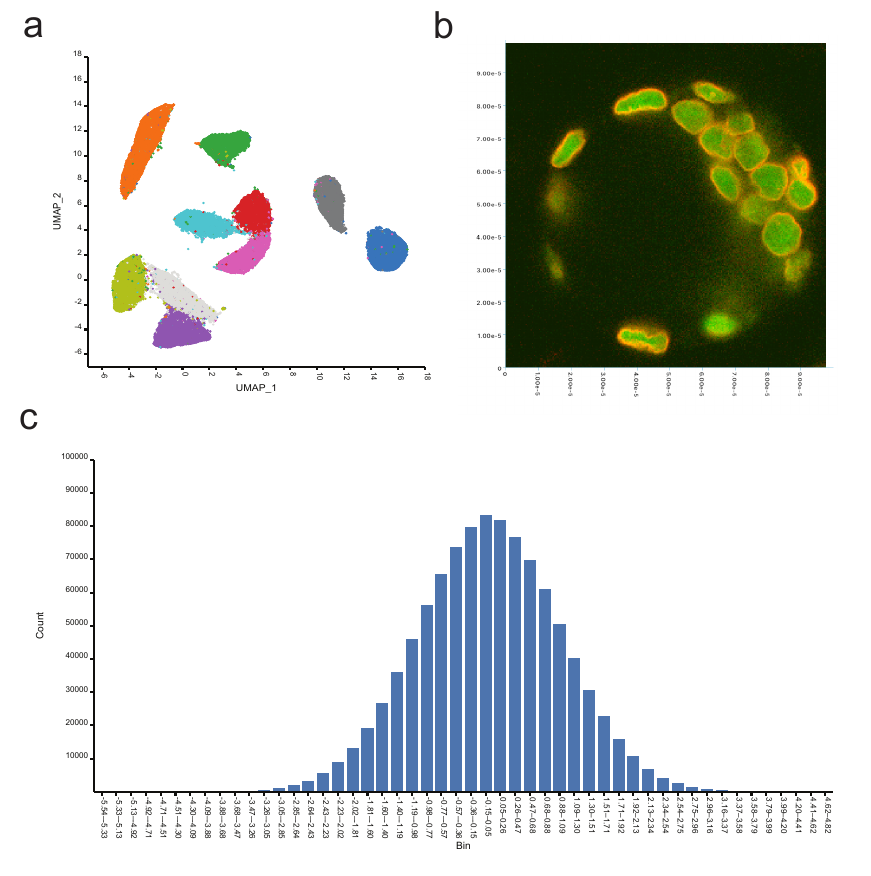}
 \caption{Examples of graphics created with Pluot. a) Scatterplot displaying a UMAP projection of the MNIST  dataset. b) Multi-scale two-channel OME-Zarr image from the Image Data Resource~\cite{moore2021omengff,williams2017image} (licensed under CC BY 4.0). c) Histogram displaying the distribution of 1 million data points.}
 \label{fig:alternative_approaches}
\end{figure}

\section{Use Cases}

We demonstrate the features of the Pluot architecture through multiple use cases.

\subsection{Scatterplot}

We demonstrate that Pluot can render millions of scatterplot points to bitmap graphics output with interactive performance, or tens of thousands of points to vector graphics output (Figure~\ref{fig:alternative_approaches}a).

\subsection{Bioimaging}

We reproduce features of Viv~\cite{manz2022viv}, demonstrating that Pluot can render multi-scale multi-channel bioimaging datasets (Figure~\ref{fig:alternative_approaches}b).
Based on the viewport extent, the appropriate resolution is chosen and the minimal set of image tiles is loaded.
Pseudocoloring is performed on the GPU.

\subsection{Histogram}

We demonstrate usage of both compute and render pipelines in our histogram implementation (Figure~\ref{fig:alternative_approaches}c).
Given an input array and number of bins, we use a GPU compute pipeline to determine the data extent and then to count the number of values for each bin.
We cache the resulting histogram values and render them using a bar plot layer.
Both the computation and rendering aspects therefore leverage GPU acceleration.

\subsection{Usage in CLIs, HTTP Servers, and GUIs}
\label{sec:cli-gui-http}

We demonstrate that Pluot can be integrated into other systems, including a command-line interface (CLI) that saves outputs to image or SVG files, an HTTP server that returns plots via byte/string responses, and a desktop application implemented using a Rust GUI framework.

\section{Discussion}

Pluot represents an architecture that is general-purpose and can be used across domains to implement custom and scalable visualization software.
The layer-based plotting paradigm that we use here can support both grammar-of-graphics-style (\ie, composing low-level layer primitives such as lines and points) or higher-level visualization authoring (\eg, using a composite layer that represents a particular chart type such as a scatterplot with X and Y axes).
Developers can implement their own layers by defining the bitmap- and vector-based drawing commands, with full control over WebGPU buffers and draw calls in the bitmap case, and with full control over SVG nodes in the vector case.
Pluot's support for asynchronous lazily-loaded data enables it to scale to dataset sizes that do not fit into memory.

Prior efforts to reuse visualization software across programming language boundaries have seen limited adoption.
This may be attributed, in part, to many prior attempts following a pattern of calling visualization rendering code implemented in interpreted languages such as JavaScript and Python, which benefit from rich ecosystems but can be slower due to their runtime overhead.
Achieving our stated design goals (Section~\ref{sec:design_goals}) required implementing a solution from scratch, enabling the reverse: calling efficient Rust/WebGPU code from other languages.
While we have implemented basic optimizations, such as caching of data arrays, we have not exhausted the space of possible optimizations.

\subsection{Limitations}

Pluot does not address all cross-platform visualization architecture problems.
We have focused on interoperability for web and desktop environments, leaving open questions regarding other environments such as mobile and extended reality.

To demonstrate the architecture, we have implemented a limited set of visual encodings
and interaction types, although these could be expanded through future work.
While we assert that this architecture is complementary to grammar-based approaches (Section~\ref{sec:related-work-grammars}), it would require further effort to use Pluot as a backend for an existing visualization grammar.

The Rust ecosystem for visualization is nascent, and does not yet contain the rich set of open-source modules present in the JavaScript, Python, or R ecosystems.
Rust and WebGPU Shading Language (WGSL) are lower-level than scripting languages such as Python or JavaScript, so unfamiliar developers may face a learning curve.
However, the 2025 Stack Overflow developer survey ranks Rust the ``most admired'' programming language, with 72.4\% of those respondents who used Rust in the prior year reporting that they want to continue using the language~\cite{stackoverflowSurvey2025}.

Targeting the web requires the developer to keep the compiled WebAssembly binary size to a minimum, as this must be loaded via the network and requires overhead to parse~\cite{pechuk2025publishtime}.
While concerns about binary size are not unique to WebAssembly, WebAssembly opens the door to importing code originally developed for non-web targets, where minimizing bundle size is not typically prioritized.
This has a cascading effect when considering transitive software dependencies.
When evaluating potential new dependencies, the effect on bundle size serves as a filter that prevents usage of certain Rust crates despite their provided features.

When using WebAssembly, web browsers currently require JavaScript ``glue'' code, complicating certain operations such as multithreading or manipulation of page elements.
Sharing memory between WebAssembly and JavaScript can also be difficult, posing obstacles when aiming to use tools from both ecosystems to operate on the same chunk of memory.
However, there are ongoing efforts by standards bodies to resolve these issues ~\cite{hunt2026whyiswebassembly}.

While leveraging GPU acceleration requires GPU hardware (either integrated or dedicated), we have designed Pluot so that CPU-based code paths can be added to support broader usage in the future, for both rendering and compute operations (albeit slower).
Alternatively, CPU-based WebGPU polyfills may emerge (analogous to SwiftShader for WebGL~\cite{githubSwiftshader}), obviating the need for CPU-specific code paths.

Pluot relies on relatively recent technologies.
Rust language version \texttt{1.0} was released in 2015, WebAssembly became available in major web browsers in 2017, and WebGPU became available in major web browsers in 2025~\cite{beaufort2025webgpuisnow}.
As software ecosystems evolve quickly, a `write once, run everywhere' solution today may not be the optimal solution for the systems of tomorrow.

\section{Conclusion}

By combining Rust, WebAssembly, and WebGPU, Pluot allows developers to write  visualization logic once and deploy it across web and desktop software via multiple programming language bindings.
This approach enables code reuse for visualization rendering, reducing redundant development efforts when building scalable, cross-platform visualization tools.
We hope that Pluot empowers both the scivis and infovis communities to develop reproducible and interoperable visualization tools that support both static and interactive use cases.

\bibliographystyle{abbrv-doi}

\bibliography{template}
\end{document}